\documentclass[12pt]{article}
\pdfoutput =1
\usepackage{float} 
\usepackage{amsmath}
\usepackage{slashed}
\usepackage{amsfonts}   
\usepackage{amssymb}

\begin{document}
\date{}
\title{{\bf{\Large Nonrelativistic pulsating strings}}}
\author{
 {\bf {\normalsize Dibakar Roychowdhury}$
$\thanks{E-mail:  dibakarphys@gmail.com, dibakarfph@iitr.ac.in}}\\
 {\normalsize  Department of Physics, Indian Institute of Technology Roorkee,}\\
  {\normalsize Roorkee 247667, Uttarakhand, India}
\\[0.3cm]
}

\maketitle
\begin{abstract}
We explore nonrelativistic (NR) pulsating string configurations over torsion Newton-Cartan (TNC) geometry having topology $ R \times S^2 $ and check the corresponding analytic integrability criteria following Kovacic's algorithm. In the first part we consider pulsating strings propagating over TNC geometry whose world-sheet theory is described by relativistic CFTs. We compute conserved charges associated with the $ 2D $ sigma model and show that the classical phase space corresponding to these NR pulsating string configurations is Liouvillian integrable. Finally, we consider nonrelativisitc scaling associated with the world-sheet d.o.f. and show that the corresponding string configuration allows even simpler integrable structure.   
\end{abstract}
\section{Overview and Motivation}
The remarkable correspondence between nonrelativistic (NR) string theory \cite{Gomis:2000bd}-\cite{Bergshoeff:2018yvt} over torsion Newton-Cartan geometry (TNC)\cite{Harmark:2017rpg}-\cite{Gallegos:2019icg} and that of Spin-Matrix Theory (SMT) limit\footnote{The SMT limit is defined as a particular NR limit near the BPS bound where one sets, $ \lambda \rightarrow 0 $, $ N= $ fixed such that $ \frac{E-J}{\lambda} =$ fixed \cite{Harmark:2018cdl}. } of \cite{Harmark:2014mpa} over $ R \times S^3 $ has recently unveiled a new exciting sector within the realm of celebrated Gauge/String duality. In a classic paper \cite{Harmark:2017rpg}, the authors show that the zero tension limit of type IIB strings propagating over the null reduced $ AdS_5 \times S^5 $ geometry essentially corresponds to SMT limit of the celebrated AdS/CFT correspondence near its unitarity bound ($ E \geq J $). Indeed the SMT limit could be thought of as being certain NR limit within $ \mathcal{N}=4 $ SYM that eventually results in a NR magnon dispersion relation \cite{Harmark:2018cdl}.

The above remarkable correspondence indeed opens up a number of exciting possibilities - for example, one might be curious to ask whether the corresponding operator spectrum in the SMT limit of $ \mathcal{N}=4 $ SYM still preserves an underlying integrable structure. In other words, how to ensure whether there exists an integrable structure on both sides of the SMT/type IIB NR string duality. One possible way to address this question is to check whether NR strings propagating over TNC geometry are integrable or not.

The purpose of this present article is to address the above issues considering the specific example of pulsating string configurations \cite{Minahan:2002rc}-\cite{Dimov:2004xi} propagating over TNC geometry with topology $ R\times S^2 $ which clearly serves as a subspace within the full null reduced $ AdS_5 \times S^5 $ target space geometry. The method that we adopt here is known as the Kovacic's algorithm \cite{kovacic1}-\cite{kovacic2} that has been found to have remarkable applications in the context of Gauge/String duality until very recently \cite{Basu:2011fw}-\cite{Nunez:2018qcj}.

We start by exploring various pulsating string configurations embedded in torsion Newton Cartan (TNC) geometry with $ R\times S^{2} $ topology \cite{Grosvenor:2017dfs}. The resulting (null reduced) nonrelativistic (NR) spacetime yields a spatial $ SO(3) $ isometry. The TNC geometry could be obtained as a result of null reduction using a combination of coordinates from the Hopf circle and the real line. In order to have a clearer understanding on the null reduction procedure as well as the corresponding coset structure one starts with Lorentzian metric of the following form,
\begin{eqnarray}
ds^2= -dt^2 +ds^2_{S^3}
\end{eqnarray}
which is a product of $ R $ times the metric on the unit three sphere embedded in $ R^{4} $ that could be thought of as being the Hopf fibration of $ S^1 \times S^2 $ \cite{Grosvenor:2017dfs},
\begin{eqnarray}
ds^2_{S^3}=\frac{1}{4}\left[ d\theta^{2}+\sin^2 \theta d\varphi^{2}+(d\psi -\cos\theta  d\varphi)^2\right] 
\end{eqnarray}
where, $ \varphi $ and $ \psi $ correspond to isometry directions whose $ U(1) $ Cartan generators are respectively, $ J_{\varphi} =\partial_{\varphi}$ and $ J_{\psi}=\partial_{\psi} $. The corresponding Killing generators satisfy two copies of $ \mathfrak{su}(2) $ algebra which is isomorphic to Lie algebra $ \mathfrak{su}(4) $.

Using the following coordinate redefinition\footnote{The coordinate $ \mathfrak{u} $ plays the role of the so called null isometry direction along the TNC manifold which we introduce next.},
\begin{eqnarray}
\mathfrak{u}=\frac{\psi}{4}-\frac{t}{2}\label{e3}
\end{eqnarray}
we arrive at the following null reduced (TNC) metric \cite{Harmark:2017rpg}-\cite{Harmark:2018cdl},
\begin{eqnarray}
ds_{TNC}^2 = 2 \tau (d \mathfrak{u}-\mathfrak{m})+\mathfrak{h}_{\mu \nu}dx^{\mu}dx^{\nu}\label{e4}
\end{eqnarray}
where we introduce the following tensor fields \cite{Grosvenor:2017dfs},
\begin{eqnarray}
\tau &=&\tau_{\mu}dx^{\mu}=\frac{1}{2}d\psi +dt-\frac{1}{2}\cos\theta d\varphi \nonumber\\
\mathfrak{m}&=&\mathfrak{m}_{\mu}dx^{\mu} =\frac{1}{4}\cos\theta d\varphi ~;~
\mathfrak{h}_{\mu \nu}dx^{\mu}dx^{\nu} =\frac{1}{4}\left[ d\theta^{2}+\sin^2 \theta d\varphi^{2}\right].
\end{eqnarray}
The above equation (\ref{e4}) is essentially the starting point of our subsequent analysis. Notice that none of the above metric coefficients (\ref{e4}) are functions of $ \mathfrak{u} $ which implies that $ \partial_{\mathfrak{u}} $ acts as a (null) Killing vector field for the full target space geometry. The resulting null reduced spacetime yields set of four Killing vectors that span the Lie algebra $ \mathfrak{su}(2)\times \mathfrak{u}(1) $. These Killing vectors $ \lbrace\xi^{\mu} \rbrace$ are what we call the generators of the so called infinitesimal diffeomorphisms associated with the reduced ($ 2+1 $)D target space geometry \cite{Grosvenor:2017dfs}, 
\begin{eqnarray}
\mathcal{L}_{\xi}\mathfrak{h}_{\mu \nu}=0~;~\mathcal{L}_{\xi}\tau_{\mu}=0~;~\mathcal{L}_{\xi}\mathfrak{m}_{\mu}=\partial_{\mu}\sigma
\end{eqnarray}
where, $ \sigma $ is the parameter associated with $ U(1) $ gauge transformation \cite{Harmark:2017rpg}.

The rest of the paper is organized as follows. We start our analysis in Section 2 by constructing the NR sigma model over TNC geometry with topology $ R \times S^2 $. The sigma model is expressed in a covariant form and the resulting theory is NR only from the perspective of the target space geometry \cite{Harmark:2017rpg}. In Section 3, we consider a particular pulsating string embedding with non zero string momentum along the null isometry direction of the TNC geometry. This geometry seems to possess an additional compact direction along which the string momentum vanishes. We allow pulsating strings to wrap around this compact direction which in turn is consistent with the fact that the string has a non zero momentum along the null isometry direction. We compute conserved charges associated with the $ 2D $ sigma model and find relation between the string oscillation number and that of the energy associated with the string. Next, we apply Kovacic's algorithm to these pulsating string configurations and establish analytic integrability by performing simple classical phase space calculations. We consider the scaling limit \cite{Harmark:2017rpg} associated with the world-sheet d.o.f. in Section 4 and find that the integrability criteria corresponding to these NR strings becomes even more trivial. Finally, we conclude in Section 5.
\section{NR sigma models on $  R \times S^2$}
We start with the relativistic closed string sigma model action,
\begin{eqnarray}
\mathcal{S}_{P}=-\frac{T}{2}\int d^{2}\sigma \sqrt{-\gamma}\gamma^{\alpha \beta}\partial_{\alpha}X^{M}\partial_{\beta}X^{N}G_{MN}=-\frac{T}{2}\int d^{2}\sigma \mathcal{L}_{P}\label{e7}
\end{eqnarray}
where, we introduce target space coordinates, $ M=(\mathfrak{u},\mu) $; $ \mu=\theta , \varphi $ together with, $ \sigma^{\alpha}=\sigma^{0},\sigma^{1} $ as being the world-sheet coordinates. Here, the coordinated $ \sigma^{1}\sim \sigma^{1}+2\pi $ is periodic.

Substituting (\ref{e4}) into (\ref{e7}) we find,
\begin{eqnarray}
\mathcal{L}_{P}=\sqrt{-\gamma}\gamma^{\alpha \beta}\varpi_{\alpha \beta}+\sqrt{-\gamma}\gamma^{\alpha \beta}\chi_{\alpha}\partial_{\beta}X^{\mathfrak{u}}\label{e8}
\end{eqnarray}
where, we introduce new functions,
\begin{eqnarray}
\chi_{\alpha}&=& 2\partial_{\alpha}t+\partial_{\alpha}\psi - \cos\theta \partial_{\alpha}\varphi \nonumber\\
\varpi_{\alpha \beta}&=&\frac{1}{4}\left( \partial_{\alpha}\theta \partial_{\beta}\theta +\partial_{\alpha}\varphi \partial_{\beta}\varphi\right) -\frac{1}{4}\cos\theta\left(\partial_{\alpha}\psi \partial_{\beta}\varphi +2\partial_{\alpha}\varphi \partial_{\beta}t \right).
\end{eqnarray}

In order to arrive at the corresponding NR string sigma model, as first step, one needs to introduce the momentum conjugate to $ X^{\mathfrak{u}} $
\begin{eqnarray}
\mathcal{P}^{\alpha}_{\mathfrak{u}}=\sqrt{-\gamma}\gamma^{\alpha \beta}\chi_{\beta}\label{e10}
\end{eqnarray}
which is conserved, $ \partial_{\alpha} \mathcal{P}^{\alpha}_{\mathfrak{u}}=0$ \cite{Harmark:2017rpg}. This stems from the fact that $ \mathcal{P}^{\alpha}_{\mathfrak{u}} $ is conserved as $ \mathfrak{u} $ being the corresponding null isometry direction associated with the target space geometry. Following \cite{Harmark:2017rpg}-\cite{Harmark:2018cdl}, our goal is to  obtain the corresponding sigma model action for closed strings in terms of remaining embedding coordinates namely, $ \tau_{\mu} $ , $ \mathfrak{m}_{\mu} $ and $ \mathfrak{h}_{\mu \nu} $. In order to do so, we focus on a sector with non zero null momentum namely, $  \gamma^{\alpha \beta}\chi_{\beta}\neq 0$ and consider dual sigma model Lagrangian of the following form \cite{Harmark:2018cdl},
\begin{eqnarray}
\tilde{\mathcal{L}}_{P}=\sqrt{-\gamma}\gamma^{\alpha \beta}\varpi_{\alpha \beta}+(\sqrt{-\gamma}\gamma^{\alpha \beta}\chi_{\alpha}-\varepsilon^{\alpha \beta}\partial_{\alpha}\zeta)\mathcal{A}_{\beta}\label{e11}
\end{eqnarray}
together with the fact, $ \varepsilon^{01}=-\varepsilon_{01}=1 $. Here, we introduce new d.o.f. on the world-sheet namely, $ \zeta $ and $ \mathcal{A}_{\alpha} $. The e.o.m. corresponding to $ \zeta $ yields, $ \varepsilon^{\alpha \beta}\partial_{\alpha}\mathcal{A}_{\beta}=0 $ which in turn implies, $ \mathcal{A}_{\alpha}=\partial_{\alpha}\vartheta $ where $ \vartheta $  is a world-sheet scalar. Therefore two sigma model descriptions (\ref{e11}) and (\ref{e8}) are equivalent (upto a total derivative) provided we identify, $ \vartheta =X^{\mathfrak{u}} $. Next, we consider variation of $ \mathcal{A}_{\alpha} $ which yields the constraint of the following form,
\begin{eqnarray}
\mathcal{P}^{\alpha}_{\mathfrak{u}}=\sqrt{-\gamma}\gamma^{  \alpha \beta}\chi_{\beta}=-\varepsilon^{\alpha \beta}\partial_{\beta}\zeta \label{e12}
\end{eqnarray}
which guarantees the conservation of world-sheet momentum off-shell. 

From (\ref{e12}), it immediately follows that in order for the string to have a non zero momentum along $ \mathfrak{u} $,
\begin{eqnarray}
\mathcal{P}_{\mathfrak{u}}=\int_{0}^{2\pi} d\sigma^{1}\mathcal{P}^{0}_{\mathfrak{u}}=\int_{0}^{2\pi} d\sigma^{1}\partial_{1}\zeta = \zeta (\sigma =0)-\zeta (\sigma =2\pi)\label{e13}
\end{eqnarray}
the world-sheet mode $ \zeta $ must have a non zero winding mode. From the perspective of the target-space geometry, these scalar modes therefore correspond to an additional compact direction along which the string winds \cite{Harmark:2018cdl}.

Next, we introduce world-sheet zweibein fields ($ \mathfrak{e}_{\alpha}~^{a} $) and their inverse,
\begin{eqnarray}
\mathfrak{e}^{\alpha}~_{a}=\frac{1}{|\mathfrak{e}|}\varepsilon^{\alpha \beta}\mathfrak{e}_{\beta}~^{b}\varepsilon_{ba}~;~|\mathfrak{e}|=\varepsilon^{\alpha \beta}\mathfrak{e}_{\alpha}~^{0}\mathfrak{e}_{\beta}~^{1}\label{e14}
\end{eqnarray}
together with the following field redefinition \cite{Harmark:2018cdl},
\begin{eqnarray}
\mathcal{A}_{\alpha}=\mathfrak{m}_{\alpha}+(\epsilon_{+}-\epsilon_{-})\mathfrak{e}_{\alpha}~^{0}+(\epsilon_{+}+\epsilon_{-})\mathfrak{e}_{\alpha}~^{1}\label{e15}
\end{eqnarray}
which upon substitution into (\ref{e11}) yields,
\begin{eqnarray}
\tilde{\mathcal{L}}_{P}=\frac{|\mathfrak{e}|}{4}\eta^{ab}\mathfrak{e}^{\alpha}~_{a}\mathfrak{e}^{\beta}~_{b}\left( \partial_{\alpha}\theta \partial_{\beta}\theta +\sin^{2}\theta\partial_{\alpha}\varphi \partial_{\beta}\varphi\right)+\frac{\varepsilon^{\alpha \beta}}{4}\cos\theta \partial_{\alpha}\varphi \partial_{\beta}\zeta \nonumber\\
+\epsilon_{+}\varepsilon^{\alpha \beta}(\chi_{\beta}+\partial_{\beta}\zeta)(\mathfrak{e}_{\alpha}~^{0}+\mathfrak{e}_{\alpha}~^{1})+\epsilon_{-}\varepsilon^{\alpha \beta}(\chi_{\beta}-\partial_{\beta}\zeta)(\mathfrak{e}_{\alpha}~^{0}-\mathfrak{e}_{\alpha}~^{1}).\label{e16}
\end{eqnarray}

Notice that in the above derivation we have used the fact,
\begin{eqnarray}
\gamma^{\alpha \beta}=\eta^{ab}\mathfrak{e}^{\alpha}~_{a}\mathfrak{e}^{\beta}~_{b}~;~\sqrt{-\gamma}=|\mathfrak{e}|
\end{eqnarray}
where, ($ a,b $) are the so called Lorentz \textit{flat} indices in two dimensions. Here, $ \epsilon_{\pm} $ are precisely the Lagrange multipliers that impose the following set of constraints,
\begin{eqnarray}
\varepsilon^{\alpha \beta}(\chi_{\beta}\pm \partial_{\beta}\zeta)(\mathfrak{e}_{\alpha}~^{0}\pm\mathfrak{e}_{\alpha}~^{1})=0
\end{eqnarray}
which could be further used to simplify (\ref{e16}).

The above action (\ref{e16}) exhibits a world-sheet Lorentz/Weyl symmetry \cite{Harmark:2018cdl},
\begin{eqnarray}
\mathfrak{e}_{\alpha}~^{0}\pm\mathfrak{e}_{\alpha}~^{1} \rightarrow \mathfrak{f}_{\pm}(\mathfrak{e}_{\alpha}~^{0}\pm\mathfrak{e}_{\alpha}~^{1})~;~\epsilon_{\pm}\rightarrow \frac{1}{\mathfrak{f}_{\pm}}
\end{eqnarray} 
which could be used to choose a gauge,
\begin{eqnarray}
\mathfrak{e}_{\alpha}~^{0}=\chi_{\alpha}~;~\mathfrak{e}_{\alpha}~^{1}=\partial_{\alpha}\zeta.\label{e20}
\end{eqnarray}

Substituting (\ref{e20}) into (\ref{e16}) and using (\ref{e14}) we finally arrive at the sigma model action for the NR strings propagating over TNC geometry with a topology $ R\times S^{2} $,
\begin{eqnarray}
\tilde{\mathcal{L}}_{NG}=\frac{\varepsilon^{\alpha \alpha'}\varepsilon^{\beta \beta'}}{\varepsilon^{\lambda\lambda'}\chi_{\lambda}\partial_{\lambda'}\zeta}(\partial_{\alpha'}\zeta \partial_{\beta'}\zeta -\chi_{\alpha'}\chi_{\beta'})\left( \partial_{\alpha}\theta \partial_{\beta}\theta +\sin^{2}\theta\partial_{\alpha}\varphi \partial_{\beta}\varphi\right)-\varepsilon^{\alpha \beta}\cos\theta \partial_{\alpha}\varphi \partial_{\beta}\zeta .\label{e21}
\end{eqnarray}

\section{Pulsating string dynamics}
\subsection{Formulation}
\subsubsection{Conserved charges}
In order to explore NR pulsating string configurations on $ R \times S^2 $, we choose to work with the following string embedding,
\begin{eqnarray}
t=t(\sigma^{0})=\kappa \sigma^{0}~;~\zeta = \mathfrak{n} ~\sigma^{1}~;~\theta = \theta (\sigma^{0})~;~\psi = \psi (\sigma^{0})~;~\varphi = \varphi (\sigma^{0})\label{e22}
\end{eqnarray}
which is consistent with the fact that the string wraps only the compact direction $ \zeta $ which results in a non-zero momentum (\ref{e13}) along the null isometry direction. Also, the above ansatz (\ref{e22}) is consistent with the fact that the string has no winding along the null isometry direction (\ref{e3}) which results in a vanishing angular momentum along the additional compact direction ($ \zeta $) \cite{Harmark:2018cdl}. 

Substituting (\ref{e22}) into (\ref{e21}) we find\footnote{We denote derivatives w.r.t. $ \sigma^{0} $ as dots.},
\begin{eqnarray}
\tilde{\mathcal{L}}_{P}=\frac{\dot{\theta}^{2}+\sin^2 \theta \dot{\varphi}^{2}}{(2\kappa +\dot{\psi}-\cos\theta \dot{\varphi})}-\cos\theta \dot{\varphi}\label{e23}
\end{eqnarray}
where, for simplicity, we set the winding number $ \mathfrak{n}=1 $.

The equations of motion that readily follow from (\ref{e23}) could be formally expressed as,
\begin{eqnarray}
\ddot{\theta}-\dot{\theta}\left( \frac{\ddot{\psi}+ \sin\theta \dot{\theta}\dot{\varphi}-\cos\theta \ddot{\varphi}}{2\kappa +\dot{\psi}-\cos\theta \dot{\varphi}}\right) -\sin\theta \cos\theta \dot{\varphi}^{2}+\frac{\sin\theta \dot{\varphi}}{2}\left( \frac{\dot{\theta}^{2}+ \sin^{2}\theta \dot{\varphi}^{2}}{2\kappa +\dot{\psi}-\cos\theta \dot{\varphi}}\right)\nonumber\\
-\frac{\sin\theta \dot{\varphi}}{2}\left(2\kappa +\dot{\psi}-\cos\theta \dot{\varphi} \right)=0, \label{e24}
\end{eqnarray}
\begin{eqnarray}
\ddot{\psi}+\sin\theta \dot{\theta}\dot{\varphi}-\cos\theta \ddot{\varphi}-\left(\frac{2\kappa +\dot{\psi}-\cos\theta \dot{\varphi}}{\dot{\theta}^{2}+ \sin^{2}\theta \dot{\varphi}^{2}} \right)\left(\dot{\theta}\ddot{\theta}+\sin\theta \cos\theta \dot{\theta}\dot{\varphi}^{2}+\sin^{2}\theta \dot{\varphi}\ddot{\varphi} \right)=0.\label{e25} 
\end{eqnarray}
The equation (\ref{e25}) corresponding to $ \psi $ could be expressed in a much simpler form by multiplying both sides of the equation by $ \frac{\dot{\theta}^{2}+ \sin^{2}\theta \dot{\varphi}^{2}}{(2\kappa +\dot{\psi}-\cos\theta \dot{\varphi})^{3}} $ and thereby expressing the L.H.S. as a total derivative in $ \sigma^{0} $ which finally yields the first order equation of the following form,
\begin{eqnarray}
 \frac{\dot{\theta}^{2}+ \sin^{2}\theta \dot{\varphi}^{2}}{(2\kappa +\dot{\psi}-\cos\theta \dot{\varphi})^{2}}=\mathcal{C}_{1}\label{e26}
\end{eqnarray}
where, $ \mathcal{C}_{1} $ stands for the constant of integration.

Finally, we note down the equation corresponding to $ \varphi $,
\begin{eqnarray}
\left( \frac{2 \sin \theta}{2\kappa +\dot{\psi}-\cos\theta \dot{\varphi}}\right) \ddot{\varphi}+\frac{4 \cos\theta \dot{\theta}\dot{\varphi}}{2\kappa +\dot{\psi}-\cos\theta \dot{\varphi}}+(1-\mathcal{C}_1)\dot{\theta}\nonumber\\
-\frac{2 \sin\theta \dot{\varphi}}{(2\kappa +\dot{\psi}-\cos\theta \dot{\varphi})^{2}}(\ddot{\psi}+\sin\theta \dot{\theta}\dot{\varphi}-\cos\theta \ddot{\varphi})=0.\label{e27}
\end{eqnarray}

A similar analysis reveals that, after multiplying by $ \sin\theta $, the equation (\ref{e27}) corresponding to $ \varphi $ could also be expressed as a total derivative in $ \sigma^{0} $ which could be integrated further to obtain,
\begin{eqnarray}
\cos\theta (1 -\mathcal{C}_{1}) -\frac{2 \sin^{2}\theta \dot{\varphi}}{2\kappa +\dot{\psi}-\cos\theta \dot{\varphi}}=\mathcal{C}_{2}
\end{eqnarray}
where, $ \mathcal{C}_{2} $ is a constant of integration. 

Next, we note down conserved charges associated with the sigma model Lagrangian (\ref{e21}). These conserved charges are precisely the energy ($ \mathcal{E} $) and the momenta ($ \mathcal{P}_{\psi} $ and $ \mathcal{P}_{\varphi} $) associated with the corresponding translations in $ t $, $ \psi $ and $ \varphi $ respectively. A straightforward computation reveals,
\begin{eqnarray}
\mathcal{E}=\frac{\partial\tilde{\mathcal{L}}_{P} }{\partial \dot{t}}=-\frac{2(\dot{\theta}^{2}+ \sin^{2}\theta \dot{\varphi}^{2})}{(2\kappa +\dot{\psi}-\cos\theta \dot{\varphi})^{2}}=-2\mathcal{C}_{1},\label{e29}
\end{eqnarray}
\begin{eqnarray}
\mathcal{P}_{\psi}=\frac{\partial\tilde{\mathcal{L}}_{P} }{\partial \dot{\psi}}=-\frac{(\dot{\theta}^{2}+ \sin^{2}\theta \dot{\varphi}^{2})}{(2\kappa +\dot{\psi}-\cos\theta \dot{\varphi})^{2}}=-\mathcal{C}_{1}=\frac{\mathcal{E}}{2},\label{e30}
\end{eqnarray}
\begin{eqnarray}
\mathcal{P}_{\varphi}=\frac{\partial\tilde{\mathcal{L}}_{P} }{\partial \dot{\varphi}}=-\cos\theta (1 -\mathcal{C}_{1}) +\frac{2 \sin^{2}\theta \dot{\varphi}}{2\kappa +\dot{\psi}-\cos\theta \dot{\varphi}}=-\mathcal{C}_{2}.\label{e31}
\end{eqnarray}
\subsubsection{Oscillation number}
In order to characterize the string dynamics, we need to find the relation between string oscillation number and that of the energy (\ref{e29}). The oscillation number is formally expressed as \cite{Beccaria:2010zn},
\begin{eqnarray}
\mathcal{N}=\frac{\sqrt{\lambda}}{2 \pi}\oint \Pi_{\theta}d\theta = \frac{\sqrt{\lambda}}{ \pi}\oint \frac{\dot{\theta}}{(2\kappa +\dot{\psi}-\cos\theta \dot{\varphi})}d\theta\label{e32}
\end{eqnarray}
where, $ \Pi_{\theta} $ is the momentum conjugate to $ \theta $ and $ \lambda $ is the t'Hooft coupling associated with the string tension ($ T $).

In order to evaluate the above integral (\ref{e32}), we first note down the Virasoro constraints (that directly follows from (\ref{e11})) associated with the NR string configuration,
\begin{eqnarray}
T_{\alpha \beta}=\mathfrak{g}_{\alpha \beta}-\frac{1}{2}\gamma_{\alpha \beta}\gamma^{\lambda \lambda'}\mathfrak{g}_{\lambda \lambda'}=0 \label{e33}
\end{eqnarray}
where we introduce, 
\begin{eqnarray}
\mathfrak{g}_{\alpha \beta}=\varpi_{\alpha \beta}+\chi_{\alpha}(\mathfrak{m}_{\beta}+\lambda_{-}\chi_{\beta}+\lambda_{+}\partial_{\beta}\zeta)
\end{eqnarray}
together with the Lagrange multipliers $ \lambda_{\pm} $.

Using conformal gauge, the above constraint (\ref{e33}) yields,
\begin{eqnarray}
\dot{\theta}^{2}+\dot{\varphi}^{2}\sin^{2}\theta +4\lambda_{-}(2\kappa+\dot{\psi}-\cos\theta \dot{\varphi})^{2}=0\label{e35}
\end{eqnarray}
which precisely matches to that with (\ref{e29}) with an identification, $ \mathcal{C}_{1}=-4\lambda_{-} $.

Substituting (\ref{e35}) into (\ref{e32}), we finally obtain
\begin{eqnarray}
\mathcal{N}= \frac{\sqrt{\lambda}}{ \pi}\sqrt{\frac{| \mathcal{E}|}{2}}\oint \sqrt{\Big|1+\frac{2 \dot{\varphi}^{2}\sin^{2}\theta}{|\mathcal{E}|(2\kappa +\dot{\psi}-\cos\theta \dot{\varphi})^2}\Big|}d\theta .\label{e36}
\end{eqnarray}

Expanding the above integral (\ref{e36}) in large string energies ($ | \mathcal{E}| \gg 1 $) we find,
\begin{eqnarray}
\mathcal{N}= \sqrt{\frac{\lambda}{2}}\sqrt{| \mathcal{E}|}+\mathcal{O}\left( \frac{1}{\sqrt{| \mathcal{E}|}}\right)
\end{eqnarray}
which therefore could be inverted to obtain,
\begin{eqnarray}
| \mathcal{E}| \approx  \frac{2 ~\mathcal{N}^2}{\lambda}.
\end{eqnarray}
\subsection{Integrability in TNC geometry}
Our goal is to understand whether the pulsating strings in TNC geometry are integrable or not. This has been a long standing problem that needs to be settled down. To address this question we adopt an analytic approach popularly known as the Kovacic's method \cite{kovacic1}-\cite{kovacic2} that has been applied with a remarkable success until very  recently \cite{Basu:2011fw}-\cite{Nunez:2018qcj}.

The basic idea behind Kovacic's algorithm is based on the usual notion of classical integrability in the context of Hamiltonian dynamics \cite{Basu:2011fw}-\cite{Stepanchuk:2012xi}. The steps that one needs to follow essentially are the following: (1) choose an invariant plane associated with the classical phase space of the dynamical system, (2) consider variations normal to this plane and obtain the corresponding normal variational equation\footnote{NVEs are typically of the form, $ y''(x)+\mathcal{P}(x)y'(x)+\mathcal{Q}(x)y(x)=0 $ where, $ \mathcal{P}(x) $ and $ \mathcal{Q}(x) $ are complex rational functions in general. Kovacic's algorithm is a prescription that essentially tells us the necessary (but not sufficient) conditions in which case the NVEs admit a Liouvillian solution  \cite{Basu:2011fw}-\cite{Stepanchuk:2012xi},\cite{Roychowdhury:2017vdo}. } (NVE) associated with these fluctuations, (3) check whether NVE admits Liouvillian solutions which essentially are simple analytic functions including algebraic functions, exponential, logarithms etc. In situations where the NVE allows a Liouvillian solution, the corresponding dynamical configuration is said to be \textit{integrable}. The cases where such solutions are not possible the correponding classical phase space configuration is termed as non-integrable.

The reader should keep a note of the fact that unlike the standard Lax pair formulation in classical mechanics, the Kovacic's algorithm does allow us to prove integrability in general. Rather it is considered to be a case by case study of different phase space configurations and to check whether a particular configuration is integrable or not. 
\subsubsection{Example 1}
In order to implement Kovacic's algorithm and to arrive at NVEs corresponding to pulsating strings in TNC geometry we first set, $\dot{\varphi}= \ddot{\varphi}=0 $ which identically satisfies (\ref{e27}) provided we also set the constant, $ \mathcal{C}_1 =1 $. Imposing this condition on the remaining two equations (\ref{e24}) and (\ref{e25}) we find,
\begin{eqnarray}
\label{ee39}
\ddot{\theta}-\frac{\dot{\theta}\ddot{\psi}}{2\kappa + \dot{\psi}}=0,\\
\ddot{\psi}-(2\kappa + \dot{\psi})\frac{\ddot{\theta}}{\dot{\theta}}=0.
\label{ee40}
\end{eqnarray}

Substituting $ \mathcal{C}_{1}=1 $ and $ \dot{\varphi}=0 $ into (\ref{e26}) we find,
\begin{eqnarray}
\frac{\dot{\theta}}{2\kappa +\dot{\psi}}=\pm 1.\label{ee41}
\end{eqnarray}

Substituting (\ref{ee41}) into (\ref{ee39}) and (\ref{ee40}) we find,
\begin{eqnarray}
\label{ee42}
\ddot{\theta}\mp\ddot{\psi}=0,\\
\ddot{\psi}\mp\ddot{\theta}=0.
\label{ee43}
\end{eqnarray}

We choose an invariant plane \cite{Roychowdhury:2017vdo} in the phase space by setting, $ \theta = 0$, $\dot{\theta}\equiv \Pi_{\theta}=0 $ which thereby yields, $ \ddot{\theta}=0 $. Substituting this into (\ref{ee43}) we find,
\begin{eqnarray}
\ddot{\psi}|_{\theta, \Pi_{\theta}\sim 0}=0.\label{e44}
\end{eqnarray}

Next, we substitute (\ref{e44}) into (\ref{ee42}) and consider fluctuations, $ \delta\theta (t) \sim \eta (t)  $ normal to this invariant plane that yields NVE in its simplest form,
\begin{eqnarray}
\ddot{\eta}(t)\approx 0
\end{eqnarray}
which admits simple algebraic (Liouvillian) solution,
\begin{eqnarray}
\eta (t)= at +b
\end{eqnarray}
thereby establishing the integrability associated with the corresponding classical phase space configuration.
\subsubsection{Example 2}
Consider a second example where we set $ \theta = \dot{\theta}=\ddot{\theta}=0 $ which trivially solves both (\ref{e24}) and (\ref{e27}). Assuming $ \frac{\theta}{\dot{\theta}}\sim 0 $ in the above limit the remaining equation (\ref{e25}) could be splitted into two parts,
\begin{eqnarray}
\label{ee47}
\ddot{\psi}-\ddot{\varphi}=0,\\
2 \kappa +\dot{\psi}-\dot{\varphi}=0.
\label{ee48}
\end{eqnarray}

Notice that, under the above assumptions we are already in a subspace $ \psi , \mathcal{P}_{\psi}=0 $ of the full dynamical phase space where $ \psi $ is arbitrary. Therefore in order to proceed further, we choose an invariant plane within this subspace by setting $ \psi =0 $ which yields, $ \dot{\psi}=0 $. Substituting this into (\ref{ee48}) we find,
\begin{eqnarray}
\varphi (t) = 2\kappa t +c.\label{ee49}
\end{eqnarray}

Substituting (\ref{ee49}) into (\ref{ee47}) and considering fluctuations, $ \delta \psi (t)\sim \eta (t) $ normal to the invariant plane in the phase space we arrive at the NVE equation,
\begin{eqnarray}
\ddot{\eta}(t)\approx 0
\end{eqnarray}
which admits a Liouvillian solution as before.
\section{A note on the scaling limit}
The string sigma model that we discuss so far is relativistic in the sense that the sigma model is described by relativistic CFTs. It is nonrelativistic only from the perspective of the target space geometry which for the present case is TNC geometry with topology $ R \times S^2 $. However, it is always possible to consider a second scaling limit \cite{Harmark:2017rpg}-\cite{Harmark:2018cdl} (associated with the world-sheet d.o.f.) that scales the vielbein fields differently which finally leads to a NR sigma model with first order time derivative. 

The purpose of this Section is therefore to take the appropriate scaling (NR) limit \cite{Harmark:2017rpg} of the world-sheet Lagrangian (\ref{e21}) and explore the corresponding pulsating string configurations. In order to do so, we consider the re-scaling of the following form, 
\begin{eqnarray}
T =\frac{\tilde{T}}{c},~\chi_{\alpha}=c^2 \partial_{\alpha} \tilde{t}+\partial_{\alpha}\tilde{\psi}-\cos \tilde{\theta}\partial_{\alpha}\tilde{\varphi},~\mathfrak{h}_{\alpha \beta}=\tilde{\mathfrak{h}}_{\alpha \beta},~\zeta = c~ \tilde{\zeta}\label{e39}
\end{eqnarray}
where we introduce a new function, $ \mathfrak{h}_{\alpha \beta}= \partial_{\alpha}\theta \partial_{\beta}\theta +\sin^{2}\theta\partial_{\alpha}\varphi \partial_{\beta}\varphi \equiv \mathfrak{h}_{\mu \nu}\partial_{\alpha}X^{\mu}\partial_{\beta}X^{\nu}$. The NR limit is achieved by taking the limit, $ c \rightarrow \infty $.

Using (\ref{e39}), we finally arrive at the NR sigma model Lagrangian,
\begin{eqnarray}
\tilde{\mathfrak{S}}_{NG}=\frac{\tilde{T}}{2}\int \tilde{\mathfrak{L}}_{NG}
\end{eqnarray}
where the Lagrangian corresponding to NR sigma model could be formally expressed as,
\begin{eqnarray}
 \tilde{\mathfrak{L}}_{NG}=\frac{\varepsilon^{\alpha \alpha'}\varepsilon^{\beta \beta'}}{\partial_{0}t}\partial_{\alpha'}t\partial_{\beta'}t ~\mathfrak{h}_{\alpha \beta}+\varepsilon^{\alpha \beta}\cos\theta \partial_{\alpha}\varphi \partial_{\beta}\zeta +\mathcal{O}(1/c^2)\label{e41}
\end{eqnarray}
where for simplicity we remove the tildes.

In order to arrive at the NR sigma model (that is first order in time derivative) we choose to work with a specific string embedding,
\begin{eqnarray}
X^{0}=t=\sigma^{0},~\varphi=\sigma^{1}+\varsigma (\sigma^{0}),~\theta = \theta (\sigma^{0})~,~\zeta = \sigma^{1}.\label{e42}
\end{eqnarray}

Substituting (\ref{e42}) into (\ref{e41}) we find,
\begin{eqnarray}
 \tilde{\mathfrak{L}}_{NG}= \sin^2\theta + \cos\theta \dot{\varsigma} 
\end{eqnarray}
which is first order in time derivative \cite{Harmark:2017rpg}. This is the Lagrangian in its simplest form. 

The equations of motion are obtained considering variations in $ \theta $ and $ \varsigma $ respectively,
\begin{eqnarray}
\cos \theta - \frac{\dot{\varsigma}}{2}& =& 0,\\
 \sin\theta \dot{\theta}& =&0.\label{e45}
\end{eqnarray}
A natural consequence of (\ref{e45}) is that $ \theta (=\theta_c) $ constant which therefore implies that, $ \varsigma (= \beta t +C )$ is a linear function in $t$.

The momentum conjugate to $ \varsigma $ could be formally expressed as, 
\begin{eqnarray}
p_{\varsigma}=\cos\theta_c
\end{eqnarray}
which is therefore a constant.

Finally, we note down the corresponding NR Hamiltonian,
\begin{eqnarray}
\tilde{\mathfrak{H}}=p_{\varsigma}\dot{\varsigma}-\tilde{\mathfrak{L}}_{NG}=|\sin^{2}\theta_{c}|=E.
\end{eqnarray}

From the above analysis, without going into the details of Kovacic's prescription, we can comment on the integrability of the corresponding classical phase space configuration. 
Notice that the phase space of the dynamical system is essentially two dimensional and the conserved charges associated with the stringy configuration is also two namely, the energy ($ E $) and the momentum ($ p_{\varsigma} $). This therefore guarantees that the NR pulsating stings propagating in TNC geometry are trivially integrable. 
\section{Summary and final remarks}
We show that the semi-classical pulsating string configurations defined over TNC geometry with topology $ R \times S^2 $ are Liouvillian intergrable in the sense of Kovacic's algorithm. The present analysis has two parts in it. In the first part we consider NR pulsating strings propagating over TNC geometry whose world-sheet theory is described by relativistic CFTs. The corresponding sigma model is found be Liouvillian integrable as the associated NVEs in the phase space admits a simple algebraic (Liouvillian) solution \cite{Basu:2011fw}. The second part of the analysis considers the so called scaling (zero tension) limit associated with the string sigma model which is related to the SMT limit of $ \mathcal{N}=4 $ SYM on $ R \times S^3 $ \cite{Harmark:2018cdl}. The corresponding NR sigma model (that is first order in time derivative and second order in spatial derivatives) is found to be trivially integrable as the number of conserved charges equals the dimension of the associated phase space.

This serves as an important as well as interesting result in the context of nonrelativistic  AdS/CFT correspondence which eventually indicates that there is a finite possibility that some of the conserved charges associated to certain specific sectors within $ \mathcal{N}=4 $ SYM might survive the SMT limit of \cite{Harmark:2014mpa} thereby preserving the underlying integrable structure in the near BPS bound. It remains to be an open question whether Kovacic's algorithm plays an useful tool in order to check integrability of NR strings in the presence of background fields. Hopefully we would be able to address this issue in the near future.\\\\ 
{\bf {Acknowledgements :}}
 The author is indebted to the authorities of IIT Roorkee for their unconditional support towards researches in basic sciences. \\\\ 


\end{document}